\documentclass[runningheads]{llncs}
\usepackage{graphicx}
\usepackage{multirow}
\usepackage{amssymb}
\usepackage{wrapfig}
\usepackage{caption}
\usepackage{floatflt}
\usepackage{amsmath}
\usepackage{hyperref}
\usepackage{xspace}
\newcommand{\name}[0]{HAMIL-QA\xspace}

\begin{document}
% TO DO: Paper title 
% Working title for registration 
%\title{Enhancing Cardiac MRI Diagnostics: Double-Module Multiple Instance Learning Approach to 3D LGE MRI Quality Assessment}
\title{\name: Hierarchical Approach to Multiple Instance Learning for Atrial LGE MRI Quality Assessment}

\titlerunning{\name}
\author{K M Arefeen Sultan$^{1,2}$ \and Md Hasibul Husain Hisham$^{1,2}$ \and Benjamin Orkild$^{1,3,4}$ \and 
Alan Morris$^{1}$ \and Eugene Kholmovski$^{5,6}$ \and 
Erik Bieging$^{5,7}$ \and Eugene Kwan$^{3,4}$ \and \\
Ravi Ranjan$^{3,4,7}$ \and Ed DiBella$^{3,5}$ \and 
Shireen Elhabian$^{1,2}$}
% index{Sultan, K M Arefeen}
% index{Hisham, Md Hasibul Husain}
% index{Orkild, Benjamin}
% index{Morris, Alan}
% index{Kholmovski, Eugene}
% index{Bieging, Erik}
% index{Kwan, Eugene}
% index{Ranjan, Ravi}
% index{DiBella, Ed}
% index{Elhabian, Shireen}

\authorrunning{K. Sultan et al.}
\institute{
Scientific Computing and Imaging Institute, University of Utah, SLC, UT \and
Kahlert School of Computing, University of Utah, SLC, UT \and
Department of Biomedical Engineering, University of Utah, SLC, UT \and
Nora Eccles Harrison Cardiovascular Research and Training Institute, University of Utah, SLC, UT \and
Department of Radiology and Imaging Sciences, University of Utah, SLC, UT \and
Department of Biomedical Engineering, Johns Hopkins, Baltimore, MD \and
Division of Cardiology, University of Utah, SLC, UT}

\maketitle              % typeset the header of the contribution
\vspace{-0.2in}\begin{abstract}

The accurate evaluation of left atrial fibrosis via high-quality 3D Late Gadolinium Enhancement (LGE) MRI is crucial for atrial fibrillation management but is hindered by factors like patient movement and imaging variability. The pursuit of automated LGE MRI quality assessment is critical for enhancing diagnostic accuracy, standardizing evaluations, and improving patient outcomes. The deep learning models aimed at automating this process face significant challenges due to the scarcity of expert annotations, high computational costs, and the need to capture subtle diagnostic details in highly variable images. This study introduces \name, a multiple instance learning (MIL) framework, designed to overcome these obstacles. \name employs a hierarchical bag and sub-bag structure that allows for targeted analysis within sub-bags and aggregates insights at the volume level. This hierarchical MIL approach reduces reliance on extensive annotations, lessens computational load, and ensures clinically relevant quality predictions by focusing on diagnostically critical image features. Our experiments show that \name surpasses existing MIL methods and traditional supervised approaches in accuracy, AUROC, and F1-Score on an LGE MRI scan dataset, demonstrating its potential as a scalable solution for LGE MRI quality assessment automation. The code is available at: \url{https://github.com/arf111/HAMIL-QA}

\keywords{Image Quality Assessment \and Weak Supervision \and Multiple Instance Learning \and Attention-based Models }

\end{abstract}
\section{Introduction}
Atrial fibrillation (AF), the most prevalent type of cardiac arrhythmia in the U.S., currently affects between 3 and 5 million individuals \cite{BAO:Col2013}. Projections suggest this number could surge to over 12 million by 2030 \cite{BAO:Col2013}. Research has established a significant connection between atrial fibrosis and the onset and recurrence of AF post-treatment \cite{BAO:Elm2015,BAO:Mar2014}. Catheter ablation, a widely adopted approach for treating AF, focuses on eradicating fibrotic tissues in the heart responsible for erratic electrical impulses by forming precise lesions or scars in these areas. This underscores the importance of accurately measuring fibrosis to effectively steer the ablation process. Despite its popularity, catheter ablation's efficacy is limited, with a recurrence rate of AF exceeding 40\% within 18 months post-procedure \cite{BAO:Ver2015}. This high recurrence rate highlights the critical need to address the limitations of current AF treatments.

Late Gadolinium Enhancement (LGE) MRI is widely utilized to quantify myocardial fibrosis and scarring. It is instrumental in assessing AF patients before catheter ablation, providing detailed insights into the atrial structure and fibrosis distribution. The geometry and fibrosis patterns identified through LGE MRI are essential for planning ablation procedures and generating patient-specific models \cite{BAO:Oak2009,BAO:Cai2021}. 
However, the quality of LGE MRI images can vary significantly, influenced by aspects such as noise, patient mobility, inconsistent breathing patterns, and suboptimal tuning of pulse sequence parameters; which can impact diagnostic precision \cite{noisylge1,noisylge2,noisylge3}.

Automating the quality assessment (QA) of LGE MRI images is critically important for clinical practices, offering to improve diagnostic accuracy, increase procedural efficiency, standardize assessments, and enhance patient outcomes. By ensuring high-quality scans for fibrosis quantification, this automation directly contributes to refining ablation strategies and guiding treatments more effectively.
%Assessing the quality of LGE MRI scans is significant because it can potentially improve the accuracy of fibrosis quantification, thereby refining ablation strategies and guiding treatment more effectively. Eliminating low-quality images from consideration allows healthcare professionals to make decisions based on more reliable imaging data. 
%
However, manual evaluation of image quality is labor-intensive and susceptible to errors, making it unsuitable for widespread application and challenging its integration into clinical practice.
%Yet, manually evaluating image quality is a tedious and error-prone task, making it an impractical solution for large-scale applications and hindering its translation to clinical routine. 
%
%The shift towards automated quality assessment could streamline processes and conserve resources, but this requires identifying image characteristics indicative of high diagnostic image quality. 
%
Automating QA can be approached naively by training a deep network that would predict a quality score from a 3D LGE volume. However, this straightforward approach faces several challenges. 
Firstly, the constraint of limited annotations, predominantly expert-driven, significantly hinders the compilation of large, annotated datasets essential for such conventional deep learning paradigms.
Secondly, the computational and memory requirements necessary for processing entire 3D volumes for image-level prediction pose a significant scalability challenge.
Moreover, the variability in LGE MRI images due to patient anatomy differences and motion artifacts complicates the development of a generalizable model. 
Lastly, ensuring that the network’s predictions are clinically relevant is a challenge, as the model must discern subtle quality nuances impacting diagnostic outcomes.
%Deep learning presents a promising avenue for automating this task, though its success heavily relies on access to many annotated images. Despite the potential benefits, the manual labeling of LGE datasets remains a demanding and slow process, often performed by experts, resulting in limited availability of labeled data for training deep learning models. This problem in turn encourages researchers to develop deep learning based models trained with limited annotations, termed as "Weakly Supervised" or "Semi-Supervised".

% [[[TODO: how MIL is applicable to our application?]]] - DONE

The quality assessment of LGE MRI scans inherently requires a weakly supervised learning approach, as typically only image-level labels are available. Multiple Instance Learning (MIL), by design, is well-suited to address this problem. 
MIL conceptualizes each 3D volume as a collection of instances (or patches), relying solely on the volume-level class label. The primary objective is to train a model capable of predicting the label for a group of instances, or ``bag". 
%The essence of MIL is to manage a collection of instances, each assigned a single class label, with the primary objective being to train a model capable of predicting the label for a group of instances, or 'bag'. 
In the context of our work, this entails the evaluation of image quality for fibrosis detection in LGE MRI scans. Each scan, considered as a \textit{bag}, comprises numerous instances, represented by hundreds of patches extracted from the scan. A bag is classified as positive if it contains at least one instance of diagnostic fibrotic tissue; otherwise, it is deemed a non-diagnostic scan. 
The choice of MIL as the foundational framework is justified by the nature of LGE MRI images, which does not guarantee the presence of diagnostic fibrotic tissue in every instance within a bag. 
Moreover, MIL optimizes computational efficiency by processing image patches rather than volumes and learns to focus on the most informative patches, thus reducing the overall processing load. 
This targeted approach aids in constructing models that are both generalizable and robust, capable of learning effectively from a limited number of annotated volumes in contrast to training networks that estimate volume-level labels from the full 3D volume. 
Furthermore, the MIL framework identifies the most diagnostic instances within a volume, aligning the learning process more closely with clinical relevance.

In this paper, we introduce \name, an MIL approach inspired by the cognitive processes employed by radiologists when assessing the diagnostic quality of LGE MRI images for the quantification of fibrosis. Although these processes may vary among radiologists, we followed the approach used by the radiologists who scored the scans in our dataset.
This approach ingeniously models the radiologist's mental strategy of evaluating scans, employing a hierarchical structure of bags and sub-bags that mimics the expert's method of systematically sweeping through the scan, slice by slice, to determine the overall quality label. 
In this framework, MIL is applied first at the sub-bag level, focusing on slices within each volume, and then at a higher bag level, integrating features learned from the sub-bag level to determine the final quality score for the entire volume. This tiered structure allows for nuanced analysis and interpretation of image data, enhancing the model's effectiveness and efficiency. 
Specifically, at the sub-bag level, the model learns to identify the most informative instances. This selective attention increases the model's ability to handle the inherent variability in LGE MRI images, as it learns to recognize and adapt to patterns across different patient anatomies and artifact influences within the sub-bags before integrating these insights at the bag level.
Experimental findings reveal that \name surpasses both traditional fully supervised models and current MIL methodologies in achieving higher accuracy, AUROC, and F1-Score metrics on a limited labeled dataset of LGE MRI scans.

%MIL's ability to facilitate learning from collections of instances without necessitating detailed annotations for each individual instance makes it exceptionally suitable for this application. 
%Our proposed methodology is tailored to address the challenges posed by limited training data in assessing LGE MRI quality, incorporating a dual-tiered approach to reduce the likelihood of misclassification at the initial stage. 

% The key contributions of this research are outlined as follows.
% \begin{enumerate}
%     \item[--] Introducing the sub-bag concept to address the challenge of limited labeled datasets in LGE MRI image quality assessment (IQA).
%     \item[--] Evaluating the effectiveness of double-tier mechanism on a limited labeled dataset.
% \end{enumerate}
\section{Related Works}

In MRI image quality control, advancements have been made to enhance the diagnostic accuracy and consistency of MRI scans. For instance, Wang et al. proposed deep generative model to enhance the quality control process of cardiac MRI segmentation, demonstrating its effectiveness across various datasets \cite{related4}. Dormont et al. proposed a framework for the automatic quality control of 3D brain T1-weighted MRI for a large clinical data warehouse \cite{related3}. K. Sultan et al. proposed a two-stage deep learning model to automate the quality assessment of LGE MRI images, crucial for evaluating left atrial fibrosis in patients with atrial fibrillation \cite{stacom}. 

Within medical image analysis, MIL has emerged as a potent framework, given its efficacy for classification tasks in histopathology. 
Quellec et al. explored MIL's applicability across various medical image and video analysis tasks, demonstrating its potential to circumvent the need for detailed pixel-level annotations \cite{related1}. One notable contribution is from Ilse et al., who proposed an attention-based deep MIL, enhancing interpretability and performance in medical image analysis by focusing on relevant instances within a bag \cite{abmil}. Furthermore, Shao et al. introduced a Dual-stream MIL Network that leverages whole slide image classification integrated with self-supervised contrastive learning, providing a nuanced approach to learning from unannotated regions within medical images effectively \cite{dsmil}. Additionally, the DTFD-MIL approach by Li et al. emphasizes a double-tier feature distillation within a MIL framework, addressing the complexities of histopathology image classification with refined feature representation and enhanced learning efficacy \cite{dtfdmil}.

To the best of our knowledge, the confluence of MIL methodologies in LGE-MRI QA remains uninvestigated. Notably, the comprehensive review by Fatima et al. on MIL underscores its versatility across a spectrum of applications, from image retrieval to disease diagnosis, reinforcing the value of integrating such methodologies within LGE-MRI QA processes \cite{related2}. 
\section{Method}

% We formulate our problem similar to a DTFD-MIL approach [Reference?] to enhance the quality control process for LGE MRI scans. 

Consider an LGE MRI volume $\mathbf{V} \in \mathbb{R}^{ R \times C \times S}$, where $R$, $C$, and $S$ represent the number of rows, columns, and slices of the volume, respectively. The diagnostic quality of the LGE MRI volume is defined as the ground truth label $Y$, which categorizes the scan as either diagnostic or non-diagnostic. 
For notational simplicity and without loss of generality, we assume each volume comprises an equal number of $M$ sub-bags, represented as $\mathcal{X} = \{\mathcal{X}_1, \mathcal{X}_2, \ldots, \mathcal{X}_M\}$, with each sub-bag $\mathcal{X}_m$ corresponding to an axial slice. A sub-bag inherits its label from its parent bag's label $Y_{GT}$. It is also assumed that each sub-bag contains a consistent number of $K$ instances, denoted by $\mathcal{X}_m = \{\mathbf{x}_m^1, \mathbf{x}_m^2, \ldots, \mathbf{x}_m^K\}$, where $\mathbf{x}_m^k \in \mathbb{R}^{r \times c}$ is a 2D image patch of $r-$rows and $c$-columns.  We randomly sample 2D patches from each axial slice to construct a sub-bag, reflecting the inherent variability within the scan. 
It is important to note that the proposed approach does not inherently require that each volume has the same number of sub-bags or that each sub-bag contains an identical number of instances, allowing for flexibility in handling diverse LGE MRI data structures.

%The axial slices of LGE MRI scans are used as the input set $X = \{X_1, X_2, \ldots, X_m\}$, where each $X_i$ represents an axial slice. 

\begin{figure}[!h]
\centering
\includegraphics[width=14cm]{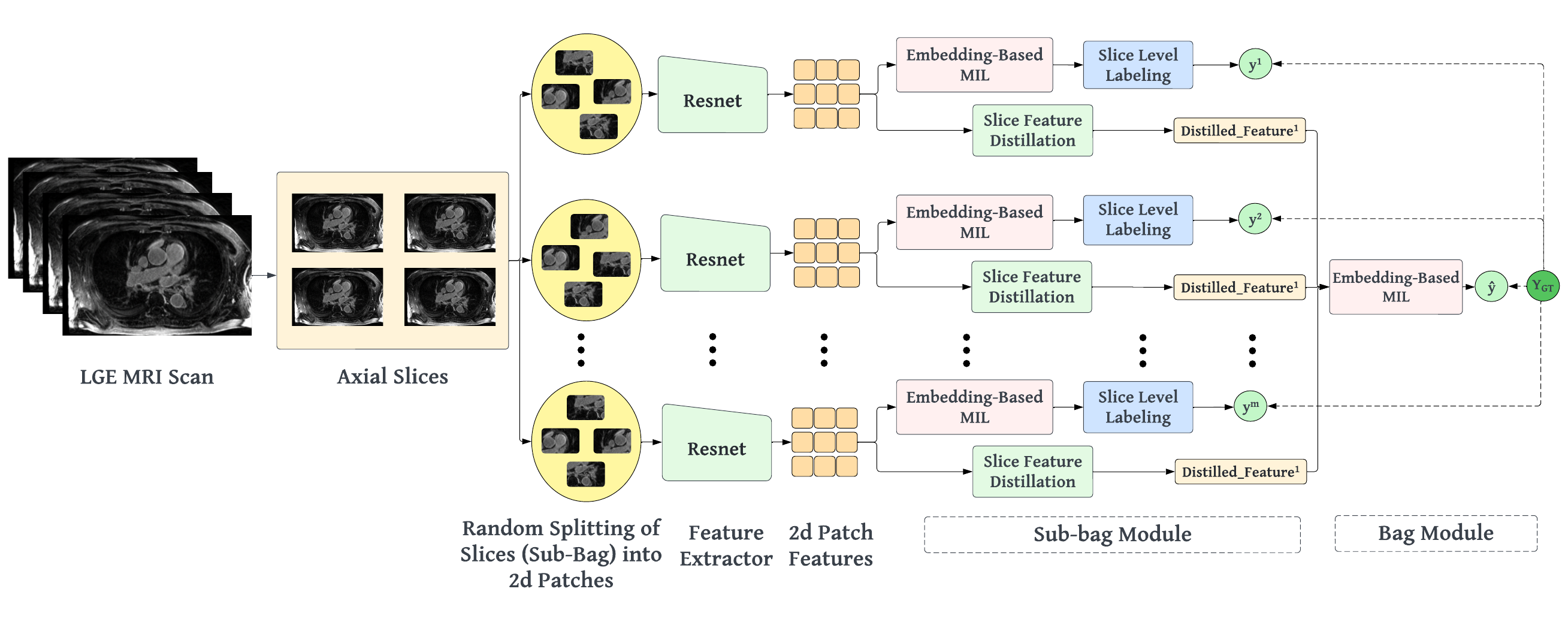}
\caption{Overview of our proposed model. For illustrative purposes, we select M axial slices (with 4 depicted as an example) at random from an LGE MRI scan, treating each slice as an individual sub-bag. We then extract random cropped patches from each of these slices. These sub-bags are initially processed by the sub-bag module. Subsequently, the output from sub-bag module is used to generate feature vectors, which are then input into the bag module. It is important to note that the ground truth label for the bags remains consistent across both sub-bag module and bag module during the training phase.}
\label{arch}
\end{figure}

We apply a transformation function $\phi(.)$ on each instance (i.e., patch) $\mathbf{x}_{m}^{k}$ to obtain the instances embedding (i.e., feature descriptors) for each sub-bag, denoted as $\mathcal{H}_m = \{\mathbf{h}_{m}^{1}, \mathbf{h}_{m}^{2}, \ldots, \mathbf{h}_{m}^{K}\}$, where $\mathbf{h}_m^k = \phi(\mathbf{x}_{m}^{k})\in \mathbb{R}^{L \times 1}$. 
We use ResNet10 to parameterize the transformation function $\phi(.)$. Then, we take a dual-stream approach on the features to perform both embedding-based MIL and slice feature distillation in our sub-bag module. 
In the first stream which is the embedding-based MIL as denoted in Figure \ref{arch}, we get the slice or sub-bag level prediction label, $\hat{y_m} = \rho_{\text{sub-bag}}(\mathcal{H}_m)$. This involves computing an attention-weighted sum of features:

\begin{equation}
\hat{y}_m = \rho_{\text{sub-bag}}(\mathcal{H}_m) = \sigma\left(\mathbf{w}_{m}\sum_{k=1}^{K} a_m^k \mathbf{h}_m^k\right),
\label{pred}
\end{equation}
% Here, $P_{Module1}$ is a function which first transforms each feature score $h_m$ weighted by the attention score $a_k$, and then maps the aggregated weighted score, $\sum_{d=1}^{D}a_{m}^{d}h_{m}^{d}$ to a single score $y_m$, which is being predictive of the diagnostic score for its corresponding sub-bag. 
where $a_m^k$ denotes the attention weight for the $k$-th instance in the $m-$th sub-bag $\mathcal{H}_m$, $\mathbf{w}_m \in \mathbb{R}^{L \times 1}$ denotes a weight vector for binary classification, and $\sigma$ represents a sigmoid activation mapping the aggregated signal to a prediction. Similar to \cite{abmil}, we calculate the attention as:
\begin{equation}
\centering
    a_m^k = \frac{\exp \left\{ \mathbf{w}^\top \tanh(\mathbf{Vh}_m^{k\top}) \right\}}
{\sum_{j=1}^K \exp \left\{ \mathbf{w}^\top \tanh(\mathbf{Vh}_m^{j\top}) \right\}}
\label{attn}
\end{equation}

Consider a dataset of $N-$volumes. The sub-bag module loss is then calculated as:

\begin{equation}
\centering
    \mathcal{L}_\text{sub-bag} = -\frac{1}{MN}\sum_{n=1}^{N}\sum_{m=1}^{M}\left[Y_n^{m} \log(y_n^{m}) + (1 - Y_n^{m}) \log(1 - y_n^{m})\right]
\end{equation}

On the other stream, slice feature distillation, we get the distilled features from transformation function $\phi(.)$ calculated as:
\begin{equation}
\centering
    \bar{\mathbf{h}} = \frac{1}{M} \sum_{n=1}^{M} \mathbf{h}_n \in \mathbb{R}^{L}
\end{equation}

%TODO: write more details here.%

In the subsequent bag module, we use an embedding based MIL from the distilled features $\bar{\mathbf{h}}$ to obtain the final LGE MRI scan level label, $\hat{y}$ by following the same procedure as equation \ref{pred} and \ref{attn}. 
The bag module loss is calculated as:
\begin{equation}
\centering
    \mathcal{L}_\text{bag} = -\frac{1}{N}\sum_{n=1}^{N}\left[Y_n \log(\hat{y}_n) + (1 - Y_n) \log(1 - \hat{y}_n)\right]
\end{equation}

Overall optimization is then:

\begin{equation}
\centering
    \theta^* = \arg \min_{\theta_1, \theta_2} ( \mathcal{L}_\text{sub-bag} + \mathcal{L}_\text{bag} )
\end{equation}

where $\theta_1$ and $\theta_2$ are the parameters of Sub-bag module and Bag module, respectively.
\section{Results}
\subsection{Dataset}
In this study, we employed a dataset comprising $424$ LGE MRI scans, each labeled for the purpose of Quality Assessment (QA) and has Left Atrium segmentations. Following the acquisition protocol detailed in \cite{BAO:Mar2014}, scans were obtained with a fine resolution of \(1.25\times1.25\times2.5 \text{mm}^3\), captured roughly 15 minutes post the administration of gadolinium using a 3D ECG-gated and respiratory-navigated gradient echo inversion recovery sequence. 
Expert reviewers rated these scans on a scale from 1 to 5. These $424$ scans have a class imbalance problem because most scans are in the $2$ to $4$ range. To address this problem, we have transformed the scores into two different labels: diagnostic and non-diagnostic. Scans with a score of $\geq 3$ are designated as diagnostic, denoted $1$, while less than $3$ is non-diagnostic and denoted as $0$.

\subsection{Data Preprocessing}
Our dataset was splitted into training and testing sets following an 80:20 split. Subsequently, the training set was further partitioned into a secondary training set and a validation set, adhering to the same 80:20 ratio. For the proposed method, we processed each scan as a series of 2D axial slices, selectively utilizing those representing the Left Atrium based on available segmentation data. In contrast with other methodologies, as detailed in Table \ref{table1}, we extracted a subvolume inclusive of the Left Atrium from the 3D scans based on the available segmentations data. We expand this subvolume by 30 units along the axial plane only. Prior to network input, all data underwent a normalization process.

\subsection{Experiments}
During training, the model is validated using AUROC across all volumes. The performance is measured by Accuracy, AUROC, and F1-Score. The details of the training are given below:

\noindent
We consider $3$ models for our comparison. Fully supervised, Attention based MIL \cite{abmil}, and DTFD-MIL framework \cite{dtfdmil}. For the fully supervised model, we only consider a 3D ResNet10 model. The 3D images are passed through the network to predict binary score, i.e., non-diagnostic and diagnostic. For ABMIL and DTFD-MIL models, we randomly pick 3D patches from the 3D images. 
All of the networks were trained for $50$ epochs using the Adam optimizer with a learning rate of $\eta = 0.0001$ and with an early stopping criteria of patience $8$. A cosine annealing learning rate scheduler was used to reduce the learning rate throughout training. The results are shown in Table \ref{table1}.
\begin{figure}[!h]
    \centering
    \includegraphics[width=7cm]{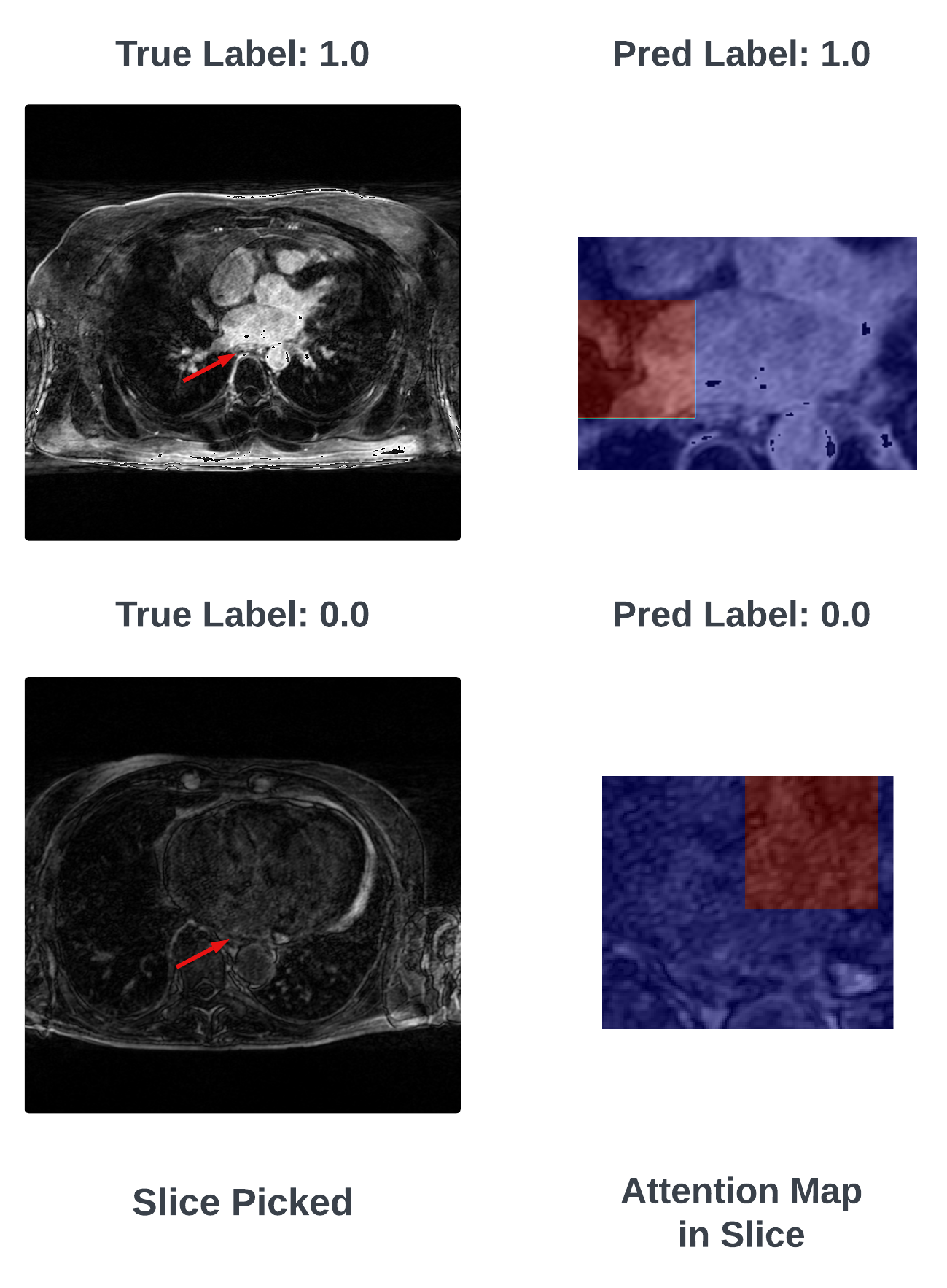}
    \caption{Heatmap of 2 scans by original LGE MRI image and by attention map of our model, respectively. In the second column, a highlighted red square marks the patch receiving the highest attention weight, with an enlarged view provided for clarity. Additionally, a red arrow on the original MRI images indicates the left atrium's position.}
    \label{fig:heatmap}
\end{figure}
\begin{table}[ht]
    \centering
    \begin{tabular}{lccc}
        \hline
        \textbf{Method} & \textbf{Acc} & \textbf{AUROC} & \textbf{F1} \\ \hline
        Fully Supervised & $0.545 \pm 0.062$ & $0.544 \pm 0.059$ & $0.419 \pm 0.210$ \\
         Classic AB-MIL \cite{abmil} & $0.643 \pm 0.028$ & $0.647 \pm 0.013$ & $0.433 \pm 0.093$ \\
         DTFD-MIL (AFS) \cite{dtfdmil} &  $0.604 \pm 0.075$ & $0.637 \pm 0.065$ & $0.242 \pm 0.242$ \\
         \name & $\mathbf{0.682 \pm 0.030}$ &  $\mathbf{0.700 \pm 0.009}$ & $\mathbf{0.596 \pm 0.084}$ \\ \hline
    \end{tabular}
    \caption{Results on our LGE MRI test set. For DTFD-MIL and our method (\name). the number of sub-bags is 6 and number of instances are 60. These numbers were determined by hyperparameter tuning. We further show the ablation experiments on these parameters in supplementary material. All of the experiments are run 3 times. The best ones are in bold.}
\label{table1}
\end{table}

For experiments, we considered different number of instances and pseudo bags to find the optimal performance. We also reported that our model is $700$x and $89$x more efficient in computation than fully supervised and two other models, respectively, since our model processes 2D image patches instead of full volume or 3D patches. The experiments are shown in the supplementary material.

To delve deeper into the efficacy of our approach, we visualized the model's attention mechanism by producing attention score heatmaps for two scans, as depicted in Figure \ref{fig:heatmap}. The visualizations reveal that our model appropriately allocates higher attention to the walls of the left atrium, which aligns with the regions commonly associated with a higher probability of fibrosis, indicating the model's capability to focus on clinically significant areas.
\section{Conclusion}
In conclusion, our study shows the development of a dual-module Multiple Instance Learning (MIL) framework is specifically designed to enhance the diagnostic quality assessment of Late Gadolinium Enhancement (LGE) Magnetic Resonance Imaging (MRI) scans. The introduction of a sub-bag concept and a double module mechanism effectively addresses the prevalent challenge of limited annotated datasets, significantly improving the model's performance metrics. The findings from this study demonstrate the framework's superior performance over traditional fully supervised and existing MIL methodologies. In summary, our study offers important contributions towards evaluating the quality of the left atrium in LGE MRI images, particularly when faced with a scarcity of labeled data.

\section*{Acknowledgements} 
This work was supported by the National Institutes of Health under grant number NHLBI-R01HL162353. The content is solely the responsibility of the authors and does not necessarily represent the official views of the National Institutes of Health.

\section*{Disclosure of Interests}
The authors have no competing interests to declare that are
relevant to the content of this article.

% \clearpage

\bibliographystyle{splncs04.bst}
\bibliography{Paper-3535.bib}

\end{document}

% TO DO: Paper title 
% Working title for registration 
%\title{Enhancing Cardiac MRI Diagnostics: Double-Module Multiple Instance Learning Approach to 3D LGE MRI Quality Assessment}
\title{Supplementary Material of \\
\name: Hierarchical Approach to Multiple Instance Learning for LGE MRI Quality Assessment}
%

% \titlerunning{\name}
\author{K M Arefeen Sultan \and Md Hasibul Husain Hisham \and Benjamin Orkild \and 
Alan Morris \and Eugene Kholmovski \and 
Erik Bieging \and Eugene Kwan \and \\
Ravi Ranjan \and Ed DiBella \and 
Shireen Elhabian}
%
% \authorrunning{Anonymous}
\institute{}

\maketitle              % typeset the header of the contribution
%

\begin{table}[]
\begin{tabular}{ccccccc}
\hline
\textbf{Method}  & \textbf{\#Patches} & \textbf{\#Sub-bags}                          & \textbf{Acc}             & \textbf{AUROC}           & \textbf{F1 Score}                              \\
\hline
% Fully Supervised & X                                  & X                               & \textbf{0.545 $\pm$ 0.062 }  & \textbf{0.544 $\pm$ 0.059 }  & \multicolumn{1}{l}{}                           \\
ABMIL            & 60                        & X                               & 0.643 $\pm$ 0.028            & 0.647 $\pm$ 0.013            & 0.433 $\pm$ 0.093           \\
DTFD-MIL         & 60                        & 5                               & 0.588 $\pm$ 0.049            & 0.568 $\pm$ 0.062            & 0.448 $\pm$ 0.110         \\
DTFD-MIL         & 60                        & 6                               & 0.604 $\pm$ 0.075            & 0.637 $\pm$ 0.065            & 0.242 $\pm$ 0.242         \\
DTFD-MIL         & 60                        & 7                               & 0.592 $\pm$ 0.037            & 0.582 $\pm$ 0.043            & 0.398 $\pm$ 0.202          \\
\name  & 60                        & 5                               & 0.647 $\pm$ 0.034            & 0.647 $\pm$ 0.034            & \textbf{0.623 $\pm$ 0.028 }          \\
\name  & 60                        & 6                               & \textbf{0.682 $\pm$ 0.030 }  & \textbf{0.700 $\pm$ 0.009 }  & 0.596 $\pm$ 0.084          \\
\name  & 60                        & 7                               & 0.663 $\pm$ 0.044            & \textbf{0.705 $\pm$ 0.010 }  & 0.524 $\pm$ 0.138          \\
\hline
ABMIL            & 80                        & X                               & 0.639 $\pm$ 0.037            & 0.637 $\pm$ 0.050            & 0.497 $\pm$ 0.130          \\
DTFD-MIL         & 80                        & 5                               & 0.553 $\pm$ 0.035            & 0.563 $\pm$ 0.047            & 0.149 $\pm$ 0.149          \\
DTFD-MIL         & 80                       & 6                               & 0.577 $\pm$ 0.058            & 0.598 $\pm$ 0.076            & 0.356 $\pm$ 0.156          \\
DTFD-MIL         & 80                        & 7                               & \textbf{0.659 $\pm$ 0.007 }  & \textbf{0.673 $\pm$ 0.016 }  & \textbf{0.669 $\pm$ 0.005 } \\
\name  & 80                        & 5                               & 0.608 $\pm$ 0.016            & 0.598 $\pm$ 0.015            & 0.504 $\pm$ 0.019           \\
\name  & 80                        & 6                               & 0.616 $\pm$ 0.041            & 0.636 $\pm$ 0.045            & 0.567 $\pm$ 0.056         \\
\name  & 80                        & 7                               & 0.651 $\pm$ 0.045            & 0.642 $\pm$ 0.045            & 0.567 $\pm$ 0.056          \\
\hline
ABMIL            & 100                       & X                               & 0.647 $\pm$ 0.020            & 0.661 $\pm$ 0.022            & 0.577 $\pm$ 0.042          \\
DTFD-MIL         & 100                       & 5                               & 0.667 $\pm$ 0.014   & 0.663 $\pm$ 0.015            & 0.601 $\pm$ 0.056           \\
DTFD-MIL         & 100                       & 6                               & 0.631 $\pm$ 0.046            & 0.616 $\pm$ 0.053            & 0.423 $\pm$ 0.187          \\
DTFD-MIL         & 100                       & 7                               & 0.647 $\pm$ 0.020            & 0.660 $\pm$ 0.021            & \textbf{0.626 $\pm$ 0.036}       \\
\name  & 100                       & 5                               & 0.620 $\pm$ 0.045            & \textbf{0.669 $\pm$ 0.020 }  & 0.458 $\pm$ 0.205          \\
\name  & 100                       & 6                               & 0.647 $\pm$ 0.045            & 0.645 $\pm$ 0.053            & 0.547 $\pm$ 0.098          \\
\name  & 100                       & 7                               & \textbf{0.675 $\pm$ 0.034 }           & 0.664 $\pm$ 0.043            & 0.513 $\pm$ 0.114         \\
\hline
\end{tabular}
\caption{Comparative analysis of Attention based MIL (ABMIL) \cite{abmil}, DTFD-MIL \cite{dtfdmil}, and our method. We used patch size (60, 60) throughout the experiments. The results are presented as mean $\pm$ standard deviation of 3 runs, reflecting the performance of each method under the specified conditions (number of patches, number of sub-bags).}
\end{table}

We have used different number of patches and number of sub-bags as hyperparameters for each method and have chosen the best model based on the validation set.

\begin{table}[h]
\centering
\begin{tabular}{|l|l|l|}
\hline
\textbf{Method} & \textbf{MACs/FLOPS} & \textbf{Number of Parameters} \\ \hline
\textbf{Fully Supervised} & 180.8B & 14.59M \\ \hline
\textbf{ABMIL} & 2.0B & 14.57M \\ \hline
\textbf{DTFD-MIL} & 1.99B & 14.57M \\ \hline
\textbf{HAMIL-QA} & 0.24B & 5.12M \\ \hline
\end{tabular}
\caption{Computational Complexity and Number of Parameters for Different Methods}
\label{tab:complexity_parameters}
\end{table}

% \clearpage

\bibliographystyle{splncs04.bst}
\bibliography{Paper-3535.bib}